\documentclass[aps,prb,twocolumn,longbibliography]{revtex4-1}
\usepackage{graphicx,color,xcolor,dcolumn,bm,amsfonts,amsmath,amssymb}
\usepackage{mathptmx,longtable,float,soul}
\usepackage{titlesec,booktabs,eqnarray}

\usepackage{xr}    
\definecolor{orange}{rgb}{1,0.5,0}
\definecolor{mypink}{rgb}{0.858, 0.188, 0.478}
\definecolor{mygray}{gray}{0.6}

\begin{document}

\preprint{v-20 / \today}
\title{Enhanced voltage-controlled magnetic anisotropy via magneto-elasticity in FePt/MgO(001) }
\author{Qurat-ul-ain$^{1}$}
\author{D. Odkhuu$^{2}$}
\author{S. H. Rhim$^{1}$}
\email[Email address: ]{sonny@ulsan.ac.kr}
\author{S. C. Hong$^{1}$}
\email[Email address: ]{schong@ulsan.ac.kr}
\affiliation{
$^{1}$Department of Physics and Energy Harvest-Storage Research Center, University of Ulsan, 
Ulsan, Republic of Korea\\
$^{2}$Department of Physics, Incheon National University, Incheon, Republic of Korea
}
\date{\today}

\begin{abstract}
The interplay between magneto-electricity (ME) and magneto-elasticity (MEL) is studied 
in the context of voltage-controlled magnetic anisotropy (VCMA).
Strain plays more than a role of changing lattice constant but that of the internal electric field in the heterostructure.
As a prototype, FePt/MgO(001) is visited, where the behavior of two interfaces are drastically different:
one exhibits switching the other does not.
Whether an external electric field ($E_{ext}$) is present or not, we found VCMA coefficient larger than 1 pJ/V$\cdot$m,
as a consequence of the rearrangement of $d$ orbitals with $m=\pm1$ and $\pm2$ 
in response to an external electric field.
In addition, magneto-crystalline anisotropy (MA) is analyzed with strain taken into account,
where non-linear feature is presented only accountable by invoking second-order MEL. 

 \end{abstract}
\maketitle

\section{Introduction}
\label{sec:intro}
The advent of spintronics has witnessed a realization of magnetic random access memory (MRAM),
which compliments or replaces conventional memories. 
This progress has relied on giant magneto-resistance (GMR)\cite{binasch1989enhanced,baibich1988giant}
and tunnel magneto-resistance (TMR)\cite{julliere1975tunneling,parkin2004giant}.
Moreover, the advancement is further pushed forward with the incorporation of 
spin-transfer torque (STT)\cite{slonczewski1996current,berger1996emission,tsoi2000generation}
and spin-orbit torque (SOT)\cite{oh2016field,qiu2014angular} for magnetization switching.
In all cases, perpendicular magneto-crystalline anisotropy (PMA) is an essential ingredient
to guarantee high bit density, lower switching current ($I_{SW}$), and thermal stability, $\Delta=KV/k_{B}T$,
where $K$ is anisotropy; $k_B$ is the Boltzmann constant; $T$ is temperature.
In spite of notable success in MRAM,
high $I_{SW}$ for switching and associated Joule heating  are major obstacles to overcome.


Magneto-electric random access memory (MeRAM) has emerged as an alternative or compliments to MRAM,
which utilizes voltage-controlled magnetic anisotropy (VCMA),
where an external electric field ($E_{ext}$) manipulates 
switching from one magnetization state to the other.
The efficiency of VCMA is characterized by a single parameter,
the VCMA coefficient, $\beta={\Delta E_{MA}}/\Delta E_{eff}$. 
The effective electric field, $E_{eff}=E_{ext}/\varepsilon_{\perp}$, where $\varepsilon_{\perp}$ is
the out-of-plane component of the dielectric tensor of an insulator, and 
$E_{MA}$ is the magneto-crystalline anisotropy energy.
In the pursuit of VCMA, various heterostructures have been explored,
where FePt/MgO is one choice.  $L1_{0}$ FePt is ferromagnetic
with a high Curie temperature of 750 K\cite{xu2014tuning} and MgO has widely been used substrate.
In addition to $E_{ext}$, 
strain can be another driving force of VCMA,
which influences $\beta$ through $\varepsilon_{\perp}$ of the insulator {\em or}
acts as an effective electric field at ferromagnetic-insulator interface even in the absence of $E_{ext}$.
Hence, comparative studies of VCMA with and without strain would be intriguing.

In this work,
magneto-electricity (ME) as well as magneto-elasticity (MEL) of FePt/MgO is investigated.
The non-linear magneto-crystalline anisotropy as a function of strain ($\eta$) is explained by invoking
second-order MEL contribution, which is usually ignored.
Fe-interface shows spin-reorientation for $4.5 < \eta < 7\%$ 
while for Pt case MA is positive regardless of $\eta$.
This difference stems from the competition
between the positive effective anisotropy and negative first-order magneto-elasticity.
Later, extremely large $\beta$ of FePt/MgO 
is presented
as a result of an interplay between $\eta$ and $E_{ext}$. 
More specifically, the rearrangement of {\em d} orbitals at the interface
in response to $E_{ext}$ is the key,
whose details are analyzed with 
band- and atom-resolved decompositions of MA.

\section{Computational methods}
\label{sec:comp-meth}
First-principles calculations have been carried out
using Vienna {\em ab initio} Simulation Package (VASP) package\cite{kresse1996efficient}
with projector augmented wave (PAW) basis\cite{blochl1994projector}.
Generalized gradient approximation is employed for the exchange-correlation potential\cite{perdew1992phys}.
Cutoff of 500 eV  for plane wave expansion and a 12$\times$12$\times$1 {\em k} mesh are used.

\begin{figure}[htbp]
\begin{center}
  \includegraphics[clip=true,width=\columnwidth]{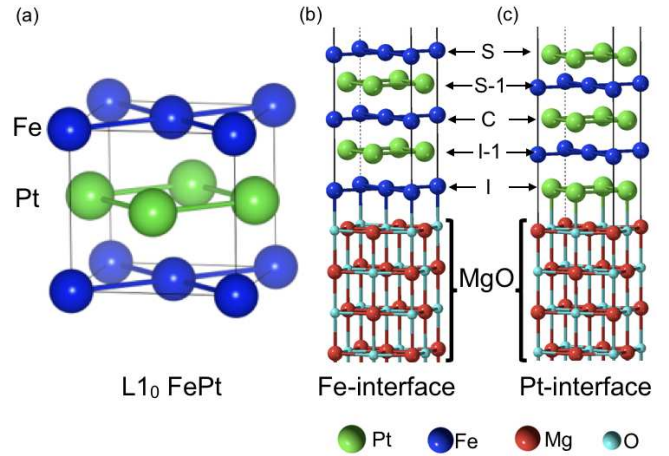}
  \caption{\label{fig:1} (a) Bulk FePt $L1_0$ structure. 5 MLs 
  of FePt on 8 MLs MgO (001) with (b) Fe- and (c) Pt-interface, respectively. Blue, green, cyan 
  and red spheres represent Fe, Pt, O and Mg atoms, respectively.
  Surface, sub-surface, center, interface, and sub-interface layers are
  denoted by $S$, $S$-1, $C$, $I$, and $I$-1 }
\end{center}
\end{figure}
Fig.~\ref{fig:1} shows the structure of bulk FePt and FePt/MgO film. 
Bulk FePt has $L1_0$ structure [Fig.~\ref{fig:1}(a)]
while the film consists of 5 monolayers (MLs) of FePt on 8 MLs MgO(001) [Fig.~\ref{fig:1}(b) and (c)].
In film, two different interfaces are taken into account
by placing (i) Fe atoms on top of O atoms [Fig.~\ref{fig:1}(b)] and (ii) Pt atoms on top of O atoms [Fig.~\ref{fig:1}(c)], which are referred to Fe- and Pt-interface, respectively. 
The vacuum region of 12 $\AA$ is taken between adjacent cells.
Both interfaces are systematically studied,
where $S$, $S$-1, $C$, $I$, and $I$-1refer to the surface, sub-surface, center, 
interface, and sub-interface layer, respectively.
The optimized lattice constant of FePt and MgO
are 3.864 and 4.212 $\AA$, respectively,
resulting in a large tensile strain ($\eta$) $\sim 8.2\%$ on the FePt layer,
assuming the MgO substrate is unstrained.
In order to study strain dependent MA of the system, $\eta$, defined as $( a - a_{FePt})/a_{FePt}$, 
is varied from $0\%$ (unstrained FePt lattice constant) to $8\%$ (nearly unstrained MgO lattice constant),
where $a_{FePt}$ is the equilibrium lattice constant of bulk FePt.
Interlayer distances are relaxed for each strain with force criteria 1$\times10^{-3}$ eV/$\AA$.
Magneto-crystalline anisotropy energy ($E_{MA}$) is determined from the total energy difference
between [100] and [001] directions,
where spin-orbit coupling (SOC) is treated in second-variational way\cite{koelling1977technique}.
Convergence of $E_{MA}$ is checked with 30$\times$30$\times$1 {\em k} mesh.
The electric field along the surface normal is applied
employing dipole layer method\cite{neugebauer1992adsorbate}.
In this work, shape anisotropy is not included in magnetic anisotropy.  
\section{Results and Discussion}
{\label{sec:results}
\begin{figure}[bp]
\centering
  \hspace{1cm}
\includegraphics[width=\columnwidth]{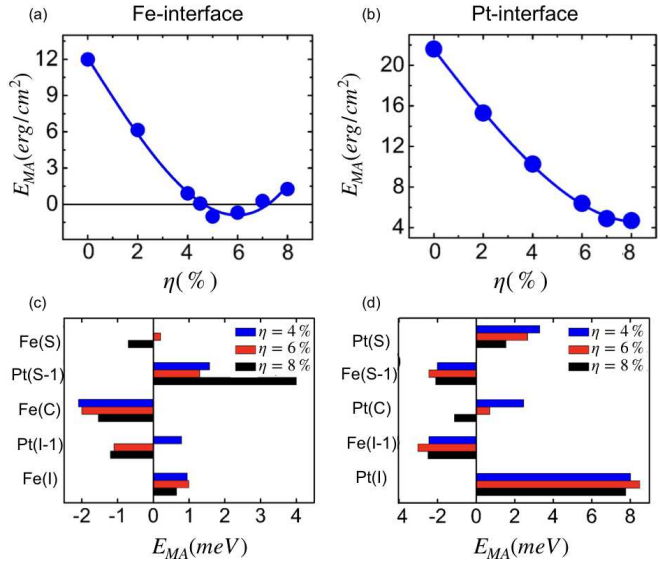}
\caption{ 
  $E_{MA}$ as a function of $\eta$ for (a) Fe- and (b) Pt-interface. 
  Circles denote calculations
  and solid line represents fitting curve according to Eq. \ref{eq:MA2}.
  Atomic layer decomposed $E_{MA}$ for (c) Fe- and (d) Pt-interface, respectively.
  Blue, red, and black bars represent $\eta = 4, 6$ and $8\%$, respectively  }
\label{fig:2}
\end{figure}
When $\eta=0\%$, $E_{MA}=12.4$ and 21.5 erg/cm$^2$ for Fe- and Pt-interface, respectively,
indicating perpendicular magnetization.
Under tensile strain, both interfaces exhibit 
parabolic curve as shown in Fig.~\ref{fig:2}(a) and (b).
However, one interface shows switching behavior but the other does not.
For Fe-interface $E_{MA}<0$ for $4.5<\eta<7\%$,
whereas for Pt-interface $E_{MA}$ decreases with strain.
The overall feature is expressed as
\begin{eqnarray}
E_{MA}= E_{MA}^0+b_1 t\sum_{k=1}^{3}\eta_k\alpha_k^2
        +\frac{1}{2}B_1 t\sum_{k=1}^{3}\eta_k^2\alpha_k^2,\label{eq:MEL}
\end{eqnarray}
where $E_{MA}^0$ is the zero strain anisotropy energy per area;
$\alpha_{k}$ and $\eta_{k}$ ($k=1,2,3$) are the direction cosines of magnetization and the strain tensor, respectively;
 $t$ is the FM film thickness;
$b_1$ and $B_1$ are the first- and second-order
MEL coefficients, respectively\cite{principles}.

MEL energy is expanded up to second-order of $\eta$,
whose coefficient $B_1$ is usually small and ignored\cite{shick1997relativistic,paes2013effective}.
However, it is explicitly taken into account here, whose consequence is discussed later.
The zero-strain anisotropy energy is approximated as 
$K_1t(1-\alpha_3^2)$ for uniaxial symmetry.
It is decomposed into bulk and interface contributions, 
$K_1 = K_1^v + K_1^i/t \approx K_1^i/t$ for thin film limit.
In tetragonal structure, $\eta_1=\eta_2=\eta$ and
the perpendicular strain $\eta_3$ is determined from magneto-elastic equation of state 
[See Supplementary Information].
Substituting the calculated strain value in Eq.~(\ref{eq:MEL}) gives
\begin{eqnarray}
\label{eq:MA2}
  E_{MA} = K_{eff}
          + \left( 1+\omega\right)b_1t\eta 
          + \left( 1-\omega\right)\frac{B_1}{2}t\eta^2,
\end{eqnarray}
where
\begin{eqnarray}
\label{eq:MA3}
  K_{eff} = K_1^i+\omega\frac{b_1^2}{c_{11}}\left(1+\frac{B_1}{2c_{11}}\right)t,  
\end{eqnarray}
and
\begin{eqnarray}
\label{eq:MA3}
  \omega = c_{11}^2/\left( c_{11}+B_1\right)^2.  
\end{eqnarray}
where $c_{11}$ is the elastic stiffness constant at constant magnetization.
The derivation of Eq.~(\ref{eq:MA2}) is also given in Supplementary Information.

Table~\ref{tab:constants} lists
magneto-elastic and effective anisotropy coefficients,
extracted by fitting {\em ab initio} results.
The second-order term, $B_1$, responsible for the non-linearity
is significantly large with 1.29 and 0.79 
$\times 10^8$ erg/cm$^3$ for Fe- and Pt- interface, respectively.
The difference in magnitudes of $B_1$ for both interfaces
arises due to different local environment of two interfaces.
Fe atoms experience larger magneto-elasticity
in the presence of MgO substrate than Pt interface.
The difference of two interfaces is further discussed now.
\begin{table}[htbp]
\centering
\caption{ 
First-order ($b_1$) and second-order ($B_1$) bulk magneto-elastic coefficients 
in ($\times 10^8$ erg/cm$^3$), and
effective anisotropy ($K_{eff}$) coefficient
in (erg/cm$^2$) for Fe- and Pt-interface, respectively.
}
\begin{ruledtabular}
  \begin{tabular}{c cc c}
Interface    & $b_1$ & $B_1$ & $K_{eff}$    \\
\hline
Fe           &  -3.16  &  1.29   & 12.44            \\
Pt           &  -2.43  &  0.79    & 21.57         
\end{tabular}
\end{ruledtabular}
\label{tab:constants}
\end{table}

The calculated $B_1$ is of the opposite sign to that of $b_1$ for both interfaces.
Further, it has been asserted that in the presence of strain, $b_1(\eta)=b_1+B_1\eta$\cite{komelj2000magnetoelastic,tian2009nonlinear}. 
In our study, the ratio $|B_1/b_1|$ is large for Fe-interface as compared to Pt-interface,
leading to a change in sign of $b_{1}$ for large strain values.
For the Fe-interface, a competition between $K_{eff}$ and $b_{1}\cdot t$
produces spin reorientation, for $4.5 <\eta <7 \%$.
On the other hand, for the Pt-interface,
$K_{eff}>b_{1}\cdot t$ results in PMA for $\eta$ up to $8\%$. 

Due to spin reorientation transition, we focus on $\eta$ = 4, 6, and 8\%.
Fig.~\ref{fig:2}(c-d) provides atomic layer resolved $E_{MA}$.
PMA mainly arises from Pt layers.
Especially, the dominant PMA contribution comes from Pt(S-1) for Fe-interface and from Pt(I)
for Pt-interface.
Pt contribution to PMA is consistent
with hard X-ray photoemission experiment\cite{ueda2016electronic}.
On the contrary, Fe atoms mostly contribute to $E_{MA}<0$,
except Fe(I) and Fe(S) layers.
Under strain, the overall behavior of $E_{MA}$ remains the same for most of the atoms with changes in magnitude only.
PMA from Fe(S), Pt(I-1) and Pt(C) at $\eta=4\%$ becomes in-plane as $\eta$ approaches to
$8\%$.

Now switching to VCMA, Fig. \ref{fig:3} shows change in MA as a function of 
$E_{eff}$ for $\eta$ = 4, 6, and 8\%. 
VCMA coefficient is defined as $\beta = \frac{\Delta E_{MA}}{\Delta E_{eff}}$
in the linear regime of $E_{eff}$ as mentioned earlier.
We choose $\epsilon_{\perp}/\epsilon_{o}$=20.0, 12.0, 9.8 for MgO when $\eta$ = 4, 6, and 8$\%$, respectively,
taken from Ref.\cite{ong2015giant}.
Large VCMA coefficients are found for both interfaces.
For Pt-interface, $\beta=$ -1.24, -1.35, and -1.36 pJ/(V$\cdot$m) under $\eta=4,$ $6,$ and $8\%$, respectively.
On the other hand, Fe-interface exhibits qualitatively different VCMA with strain.
The V-shape curve is apparent for $\eta=4$ and $6\%$ with $\beta$ = 1.70 (-0.44) and 0.79 (-1.53) 
when $E_{eff} >0$ ($E_{eff}< 0$), respectively.
At $\eta=8\%$, the VCMA curve changes to $\Lambda$-shape 
with $\beta=$ -1.77 (1.68) under $E_{eff} > 0$ ($E_{eff} < 0$).

\begin{figure}[htbp]
\centering
  \hspace{1cm}
\includegraphics[width=\columnwidth]{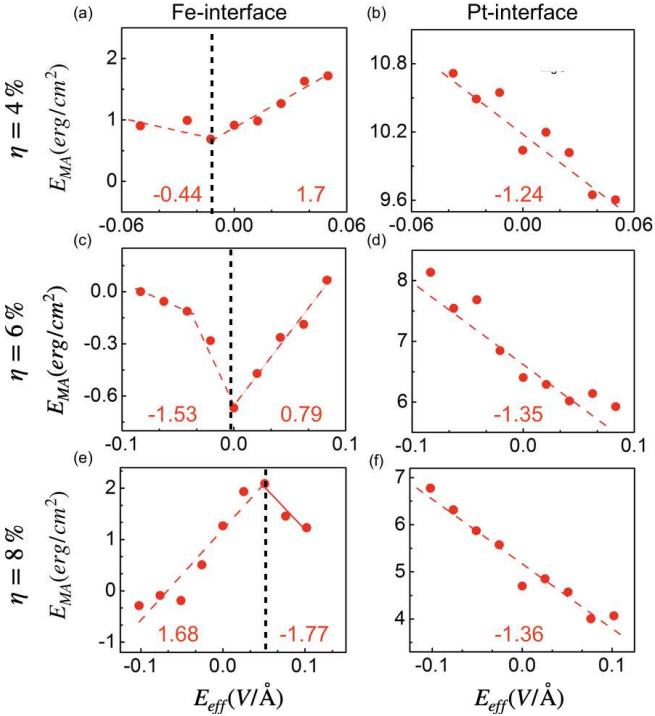}
  \caption{
VCMA of FePt/MgO heterostructure 
at different strain values for Fe- (left-panel) and Pt-interface (right-panel), respectively.
Upper, middle, and lower row represent strain ($\eta$) of 4, 6, and 8\%, respectively.
VCMA coefficient are denoted inside each plot.} 
\label{fig:3}
\end{figure}
\begin{figure}[htbp]
\centering
  \hspace{1cm}
\includegraphics[width=\columnwidth]{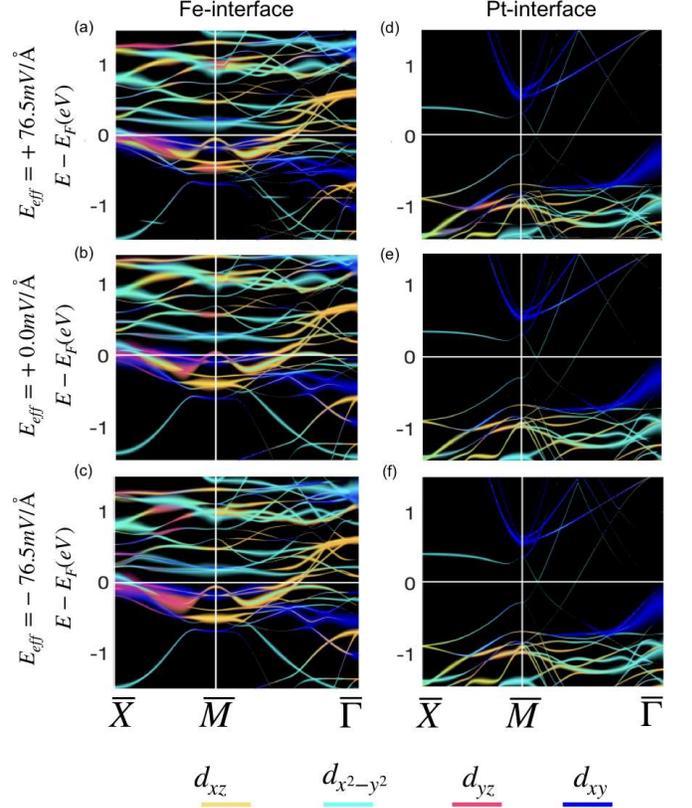}
  \caption{
Orbital resolved interfacial (a-c) Fe {\em d} bands for minority spin,
(d-f) Pt {\em d} bands for majority spin along $\overline X$-$\overline M$-$\overline\Gamma$ at $\eta=8\%$
under $E_{eff}$= +76.5 mV/$\AA$, 0, -76.5 mV/$\AA$.
Blue, cyan, pink, and yellow for {\em d$_{xy}$}, {\em d$_{x^2-y^2}$},
{\em d$_{yz}$}, and {\em d$_{xz}$}. The {\em d$_{z^2}$} bands can contribute negatively to PMA and are not plotted here. }
\label{fig:4}
\end{figure}

To understand the underlying mechanism of strain-induced MA and VCMA,
orbital resolved bands at $\eta=8\%$ are plotted in Fig.~\ref{fig:4} along high symmetry lines
in two-dimensional Brillouin zone (BZ) under $E_{eff}$= +76.5, 0, and -76.5 mV/$\AA$. 
The $\eta=8\%$ case are discussed in detail as it shows largest VCMA coefficient. 
For Fe- and Pt-interfaces, only the minority spin channel of Fe {\em d} bands 
and majority spin channel of Pt {\em d} bands are presented, respectively,
as other spin channels do not contribute significantly to PMA. 
The {\em d$_{z^2}$} orbitals for both interfaces can contribute negatively to PMA
and are shown in Supplementary Information.
Both spin channels for Fe and Pt {\em d} bands at $\eta=8\%$, $6\%$, and $4\%$ 
are also provided in Fig. S1 and Fig. S2 in Supplementary Information, respectively.

In the framework of perturbation theory\cite{wang1993first},
positive (negative) $E_{MA}$ comes from spin-orbit coupling (SOC)
between the unoccupied and occupied majority or minority spin states
with the same (different) magnetic quantum number through $\ell_z (\ell_x)$.
This approach has been widely applied in various systems
\cite{kohji10:prb,odkhuu2011engineering,odkhuu13:prb,hotta13:prl,odkhuu2016126,qurat201869}.

First, we discuss without $E_{ext}$, namely, strain-induced MA. 
For Fe-interface,
$E_{MA}>0$ arises from $\langle d_{xy} \downarrow|\ell_z|d_{x^2-y^2}\downarrow \rangle$ 
and   
$\langle d_{yz} \downarrow|\ell_z|d_{xz}\downarrow \rangle$
along $\overline{XM}$ [Fig.~\ref{fig:4}(b)].
Similarly, for Pt- interface $E_{MA}>0$ mainly comes from
$\langle d_{x^2-y^2} \uparrow|\ell_z|d_{xy}\uparrow \rangle$
along $\overline{M\Gamma}$ [Fig.~\ref{fig:4}(e)].
As tensile strain decreases, $d$ bands experience
overall downward shift for Fe-interface.
However, for Pt-interface, $d_{xy}$ and 
$d_{x^2-y^2}$ moves upward and downward, respectively, with decreasing strain,
which is shown in Fig.~\ref{fig:S1} and Fig.~\ref{fig:S2} in Supplementary Information. 
Strain driven band rearrangement leads to substantial
change in $E_{MA}$ as $E_{MA} \propto \Delta =1/(e_u - e_o)$, where
$e_u$ ($e_o$) denotes energies of unoccupied (occupied) bands.
In particular, at $\eta = 6\%$ for Fe-interface,
$E_{MA}<0$ comes from 
$\langle d_{yz} \downarrow|\ell_x|d_{xy}\downarrow \rangle$ around $\frac{1}{2}\overline {XM}$. 
Also, at $\eta = 8\%$, $E_{MA} >0$ is through
$\langle d_{xy} \downarrow|\ell_z|d_{x^2-y^2}\downarrow \rangle$ around $\overline X$. 

Moving to VCMA, bands shift at $\eta=8\%$ under $E_{eff}= \pm 76.5$ $mV/\AA$
are shown in top and bottom panels of Fig.~\ref{fig:4}.
To understand in a simple picture, a schematic diagram is illustrated in Fig.~\ref{fig:5}.
$\Delta^{\alpha}=1/\left(e_u-e_o\right)$ ($\alpha=+,0,-$) denotes
the inverse of the energy difference between unoccupied and occupied bands
when $E_{eff}$ is positive, zero, and negative, respectively.

Summing all SOC matrices, $\Delta^0> \Delta^+> \Delta^-$ justifies the $\Lambda$-shaped VCMA for Fe-interface. 
Under zero-field, occupied {\em d$_{x^2-y^2}$} ({\em d$_{xz}$}) bands
couples with unoccupied {\em d$_{xy}$} ({\em d$_{yz}$}) bands at $\frac{1}{2}\overline {XM}$, giving $E_{MA}>0$.
With $E_{eff}=\pm$ 76.5 mV/$\AA$, unoccupied bands {\em d$_{xy}$} and {\em d$_{yz}$} becomes occupied, resulting in $E_{MA}=0$.
Moreover, when $E_{eff} >0$, {\em d$_{xy}$} and {\em d$_{xz}$} occupied bands along with
{\em d$_{x^2-y^2}$} and {\em d$_{yz}$} unoccupied
bands move towards $E_F$  at $\overline X$ and $\overline M$, providing large PMA.
While when $E_{eff} <0$, these bands 
moves away from $E_F$,  as a result contributing small PMA.
On the other hand, for Pt-interface, $\Delta^-> \Delta^0> \Delta^+$ explains linear VCMA.
When $E_{eff} <0$, the unoccupied  {\em d$_{xy}$} band and occupied {\em d$_{x^2-y^2}$} band at $\overline X$, shift towards 
$E_F$ with respect to zero-field, resulting in enhanced PMA.
However, when $E_{eff} >0$, both these bands move away from $E_F$ as compared to zero-field, hence PMA is reduced.

\begin{figure}[bp]
\centering
  \hspace{1cm}
\includegraphics[width=\columnwidth]{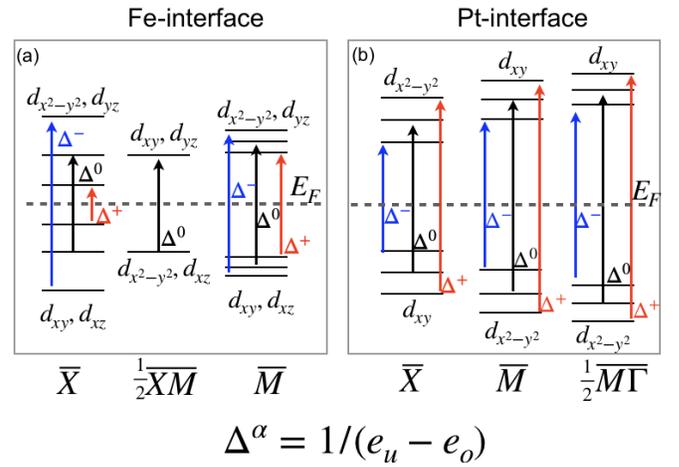}
\caption{Schematic diagram of bands shift under $E_{eff}$.
$\Delta^{\alpha}=\frac{1}{e_u - e_o}$ represents the strength of SOC,
where $e_u$ ($e_o$) are energies of unoccupied (occupied) band;
$\alpha=+,0,-$ denotes when $E_{eff} > 0 $, $E_{eff}=0$, and $E_{eff}<0$, respectively.
Vertical arrows indicates possible coupling responsible for PMA.}
\label{fig:5}
\end{figure}
\section{Conclusions}
\label{sec:conclusions}
In summary, we investigated strain dependent voltage-controlled magnetic anisotropy for both Fe- and Pt-interfaces of FePt/MgO(001) film using {\em ab initio} electronic structure calculations.
We predicted a huge VCMA coefficient $\sim$ 1.77 pJ/(V$\cdot$m) due to the internal electric field as a result of strain.
Moreover, magneto-crystalline anisotropy as a function of strain is also discussed.
The strain-dependent non-linear magneto-crystalline anisotropy is explained by invoking second-order magneto-elastic (MEL)
term in MA energy.
Fe-interface shows spin-reorientation for $4.5 < \eta < 7\%$ as a consequence of the competition 
between the positive $K_{eff}$ and negative $b_1\cdot t$. 
Magneto-crystalline anisotropy turns out to be extremely sensitive to strain and interface.
Our finding provides a direction for experiments to achieve enhanced VCMA coefficient along with large PMA
for ultra-low power nonvolatile memory devices.

\begin{acknowledgments}
  This work was supported by National Research Foundation of Korea (NRF) grant
  (NRF-2018R1A4A1020696 and NRF-2019R1I1A3A01059880).
\end{acknowledgments}

\begin{thebibliography}{29}%
\makeatletter
\providecommand \@ifxundefined [1]{%
 \@ifx{#1\undefined}
}%
\providecommand \@ifnum [1]{%
 \ifnum #1\expandafter \@firstoftwo
 \else \expandafter \@secondoftwo
 \fi
}%
\providecommand \@ifx [1]{%
 \ifx #1\expandafter \@firstoftwo
 \else \expandafter \@secondoftwo
 \fi
}%
\providecommand \natexlab [1]{#1}%
\providecommand \enquote  [1]{``#1''}%
\providecommand \bibnamefont  [1]{#1}%
\providecommand \bibfnamefont [1]{#1}%
\providecommand \citenamefont [1]{#1}%
\providecommand \href@noop [0]{\@secondoftwo}%
\providecommand \href [0]{\begingroup \@sanitize@url \@href}%
\providecommand \@href[1]{\@@startlink{#1}\@@href}%
\providecommand \@@href[1]{\endgroup#1\@@endlink}%
\providecommand \@sanitize@url [0]{\catcode `\\12\catcode `\$12\catcode
  `\&12\catcode `\#12\catcode `\^12\catcode `\_12\catcode `\%12\relax}%
\providecommand \@@startlink[1]{}%
\providecommand \@@endlink[0]{}%
\providecommand \url  [0]{\begingroup\@sanitize@url \@url }%
\providecommand \@url [1]{\endgroup\@href {#1}{\urlprefix }}%
\providecommand \urlprefix  [0]{URL }%
\providecommand \Eprint [0]{\href }%
\providecommand \doibase [0]{http://dx.doi.org/}%
\providecommand \selectlanguage [0]{\@gobble}%
\providecommand \bibinfo  [0]{\@secondoftwo}%
\providecommand \bibfield  [0]{\@secondoftwo}%
\providecommand \translation [1]{[#1]}%
\providecommand \BibitemOpen [0]{}%
\providecommand \bibitemStop [0]{}%
\providecommand \bibitemNoStop [0]{.\EOS\space}%
\providecommand \EOS [0]{\spacefactor3000\relax}%
\providecommand \BibitemShut  [1]{\csname bibitem#1\endcsname}%
\let\auto@bib@innerbib\@empty
\bibitem [{\citenamefont {Binasch}\ \emph {et~al.}(1989)\citenamefont
  {Binasch}, \citenamefont {Gr{\"u}nberg}, \citenamefont {Saurenbach},\ and\
  \citenamefont {Zinn}}]{binasch1989enhanced}%
  \BibitemOpen
  \bibfield  {author} {\bibinfo {author} {\bibfnamefont {G.}~\bibnamefont
  {Binasch}}, \bibinfo {author} {\bibfnamefont {P.}~\bibnamefont
  {Gr{\"u}nberg}}, \bibinfo {author} {\bibfnamefont {F.}~\bibnamefont
  {Saurenbach}}, \ and\ \bibinfo {author} {\bibfnamefont {W.}~\bibnamefont
  {Zinn}},\ }\href@noop {} {\bibfield  {journal} {\bibinfo  {journal} {Phys.
  Rev. B}\ }\textbf {\bibinfo {volume} {39}},\ \bibinfo {pages} {4828}
  (\bibinfo {year} {1989})}\BibitemShut {NoStop}%
\bibitem [{\citenamefont {Baibich}\ \emph {et~al.}(1988)\citenamefont
  {Baibich}, \citenamefont {Broto}, \citenamefont {Fert}, \citenamefont
  {Van~Dau}, \citenamefont {Petroff}, \citenamefont {Etienne}, \citenamefont
  {Creuzet}, \citenamefont {Friederich},\ and\ \citenamefont
  {Chazelas}}]{baibich1988giant}%
  \BibitemOpen
  \bibfield  {author} {\bibinfo {author} {\bibfnamefont {M.~N.}\ \bibnamefont
  {Baibich}}, \bibinfo {author} {\bibfnamefont {J.~M.}\ \bibnamefont {Broto}},
  \bibinfo {author} {\bibfnamefont {A.}~\bibnamefont {Fert}}, \bibinfo {author}
  {\bibfnamefont {F.~N.}\ \bibnamefont {Van~Dau}}, \bibinfo {author}
  {\bibfnamefont {F.}~\bibnamefont {Petroff}}, \bibinfo {author} {\bibfnamefont
  {P.}~\bibnamefont {Etienne}}, \bibinfo {author} {\bibfnamefont
  {G.}~\bibnamefont {Creuzet}}, \bibinfo {author} {\bibfnamefont
  {A.}~\bibnamefont {Friederich}}, \ and\ \bibinfo {author} {\bibfnamefont
  {J.}~\bibnamefont {Chazelas}},\ }\href@noop {} {\bibfield  {journal}
  {\bibinfo  {journal} {Phys. Rev. Lett.}\ }\textbf {\bibinfo {volume} {61}},\
  \bibinfo {pages} {2472} (\bibinfo {year} {1988})}\BibitemShut {NoStop}%
\bibitem [{\citenamefont {Julliere}(1975)}]{julliere1975tunneling}%
  \BibitemOpen
  \bibfield  {author} {\bibinfo {author} {\bibfnamefont {M.}~\bibnamefont
  {Julliere}},\ }\href@noop {} {\bibfield  {journal} {\bibinfo  {journal}
  {Phys. Lett. A}\ }\textbf {\bibinfo {volume} {54}},\ \bibinfo {pages} {225}
  (\bibinfo {year} {1975})}\BibitemShut {NoStop}%
\bibitem [{\citenamefont {Parkin}\ \emph {et~al.}(2004)\citenamefont {Parkin},
  \citenamefont {Kaiser}, \citenamefont {Panchula}, \citenamefont {Rice},
  \citenamefont {Hughes}, \citenamefont {Samant},\ and\ \citenamefont
  {Yang}}]{parkin2004giant}%
  \BibitemOpen
  \bibfield  {author} {\bibinfo {author} {\bibfnamefont {S.~S.}\ \bibnamefont
  {Parkin}}, \bibinfo {author} {\bibfnamefont {C.}~\bibnamefont {Kaiser}},
  \bibinfo {author} {\bibfnamefont {A.}~\bibnamefont {Panchula}}, \bibinfo
  {author} {\bibfnamefont {P.~M.}\ \bibnamefont {Rice}}, \bibinfo {author}
  {\bibfnamefont {B.}~\bibnamefont {Hughes}}, \bibinfo {author} {\bibfnamefont
  {M.}~\bibnamefont {Samant}}, \ and\ \bibinfo {author} {\bibfnamefont {S.-H.}\
  \bibnamefont {Yang}},\ }\href@noop {} {\bibfield  {journal} {\bibinfo
  {journal} {Nat. Mater.}\ }\textbf {\bibinfo {volume} {3}},\ \bibinfo {pages}
  {862} (\bibinfo {year} {2004})}\BibitemShut {NoStop}%
\bibitem [{\citenamefont {Slonczewski}(1996)}]{slonczewski1996current}%
  \BibitemOpen
  \bibfield  {author} {\bibinfo {author} {\bibfnamefont {J.~C.}\ \bibnamefont
  {Slonczewski}},\ }\href@noop {} {\bibfield  {journal} {\bibinfo  {journal}
  {J. Magn. Magn. Mater.}\ }\textbf {\bibinfo {volume} {159}},\ \bibinfo
  {pages} {L1} (\bibinfo {year} {1996})}\BibitemShut {NoStop}%
\bibitem [{\citenamefont {Berger}(1996)}]{berger1996emission}%
  \BibitemOpen
  \bibfield  {author} {\bibinfo {author} {\bibfnamefont {L.}~\bibnamefont
  {Berger}},\ }\href@noop {} {\bibfield  {journal} {\bibinfo  {journal} {Phys.
  Rev. B}\ }\textbf {\bibinfo {volume} {54}},\ \bibinfo {pages} {9353}
  (\bibinfo {year} {1996})}\BibitemShut {NoStop}%
\bibitem [{\citenamefont {Tsoi}\ \emph {et~al.}(2000)\citenamefont {Tsoi},
  \citenamefont {Jansen}, \citenamefont {Bass}, \citenamefont {Chiang},
  \citenamefont {Tsoi},\ and\ \citenamefont {Wyder}}]{tsoi2000generation}%
  \BibitemOpen
  \bibfield  {author} {\bibinfo {author} {\bibfnamefont {M.}~\bibnamefont
  {Tsoi}}, \bibinfo {author} {\bibfnamefont {A.}~\bibnamefont {Jansen}},
  \bibinfo {author} {\bibfnamefont {J.}~\bibnamefont {Bass}}, \bibinfo {author}
  {\bibfnamefont {W.-C.}\ \bibnamefont {Chiang}}, \bibinfo {author}
  {\bibfnamefont {V.}~\bibnamefont {Tsoi}}, \ and\ \bibinfo {author}
  {\bibfnamefont {P.}~\bibnamefont {Wyder}},\ }\href@noop {} {\bibfield
  {journal} {\bibinfo  {journal} {Nature}\ }\textbf {\bibinfo {volume} {406}},\
  \bibinfo {pages} {46} (\bibinfo {year} {2000})}\BibitemShut {NoStop}%
\bibitem [{\citenamefont {Oh}\ \emph {et~al.}(2016)\citenamefont {Oh},
  \citenamefont {Baek}, \citenamefont {Kim}, \citenamefont {Lee}, \citenamefont
  {Lee}, \citenamefont {Yang}, \citenamefont {Park}, \citenamefont {Lee},
  \citenamefont {Kim}, \citenamefont {Go} \emph {et~al.}}]{oh2016field}%
  \BibitemOpen
  \bibfield  {author} {\bibinfo {author} {\bibfnamefont {Y.-W.}\ \bibnamefont
  {Oh}}, \bibinfo {author} {\bibfnamefont {S.-h.~C.}\ \bibnamefont {Baek}},
  \bibinfo {author} {\bibfnamefont {Y.~M.}\ \bibnamefont {Kim}}, \bibinfo
  {author} {\bibfnamefont {H.~Y.}\ \bibnamefont {Lee}}, \bibinfo {author}
  {\bibfnamefont {K.-D.}\ \bibnamefont {Lee}}, \bibinfo {author} {\bibfnamefont
  {C.-G.}\ \bibnamefont {Yang}}, \bibinfo {author} {\bibfnamefont {E.-S.}\
  \bibnamefont {Park}}, \bibinfo {author} {\bibfnamefont {K.-S.}\ \bibnamefont
  {Lee}}, \bibinfo {author} {\bibfnamefont {K.-W.}\ \bibnamefont {Kim}},
  \bibinfo {author} {\bibfnamefont {G.}~\bibnamefont {Go}},  \emph {et~al.},\
  }\href@noop {} {\bibfield  {journal} {\bibinfo  {journal} {Nat.
  Nanotechnol.}\ }\textbf {\bibinfo {volume} {11}},\ \bibinfo {pages} {878}
  (\bibinfo {year} {2016})}\BibitemShut {NoStop}%
\bibitem [{\citenamefont {Qiu}\ \emph {et~al.}(2014)\citenamefont {Qiu},
  \citenamefont {Deorani}, \citenamefont {Narayanapillai}, \citenamefont {Lee},
  \citenamefont {Lee}, \citenamefont {Lee},\ and\ \citenamefont
  {Yang}}]{qiu2014angular}%
  \BibitemOpen
  \bibfield  {author} {\bibinfo {author} {\bibfnamefont {X.}~\bibnamefont
  {Qiu}}, \bibinfo {author} {\bibfnamefont {P.}~\bibnamefont {Deorani}},
  \bibinfo {author} {\bibfnamefont {K.}~\bibnamefont {Narayanapillai}},
  \bibinfo {author} {\bibfnamefont {K.-S.}\ \bibnamefont {Lee}}, \bibinfo
  {author} {\bibfnamefont {K.-J.}\ \bibnamefont {Lee}}, \bibinfo {author}
  {\bibfnamefont {H.-W.}\ \bibnamefont {Lee}}, \ and\ \bibinfo {author}
  {\bibfnamefont {H.}~\bibnamefont {Yang}},\ }\href@noop {} {\bibfield
  {journal} {\bibinfo  {journal} {Sci. Rep.}\ }\textbf {\bibinfo {volume}
  {4}},\ \bibinfo {pages} {4491} (\bibinfo {year} {2014})}\BibitemShut
  {NoStop}%
\bibitem [{\citenamefont {Xu}\ \emph {et~al.}(2014)\citenamefont {Xu},
  \citenamefont {Sun}, \citenamefont {Chen}, \citenamefont {Zhou},
  \citenamefont {Heald}, \citenamefont {Bergman}, \citenamefont {Sanyal},\ and\
  \citenamefont {Chow}}]{xu2014tuning}%
  \BibitemOpen
  \bibfield  {author} {\bibinfo {author} {\bibfnamefont {D.}~\bibnamefont
  {Xu}}, \bibinfo {author} {\bibfnamefont {C.-J.}\ \bibnamefont {Sun}},
  \bibinfo {author} {\bibfnamefont {J.-S.}\ \bibnamefont {Chen}}, \bibinfo
  {author} {\bibfnamefont {T.-J.}\ \bibnamefont {Zhou}}, \bibinfo {author}
  {\bibfnamefont {S.~M.}\ \bibnamefont {Heald}}, \bibinfo {author}
  {\bibfnamefont {A.}~\bibnamefont {Bergman}}, \bibinfo {author} {\bibfnamefont
  {B.}~\bibnamefont {Sanyal}}, \ and\ \bibinfo {author} {\bibfnamefont {G.~M.}\
  \bibnamefont {Chow}},\ }\href@noop {} {\bibfield  {journal} {\bibinfo
  {journal} {J. Appl. Phys.}\ }\textbf {\bibinfo {volume} {116}},\ \bibinfo
  {pages} {143902} (\bibinfo {year} {2014})}\BibitemShut {NoStop}%
\bibitem [{\citenamefont {Kresse}\ and\ \citenamefont
  {Furthm{\"u}ller}(1996)}]{kresse1996efficient}%
  \BibitemOpen
  \bibfield  {author} {\bibinfo {author} {\bibfnamefont {G.}~\bibnamefont
  {Kresse}}\ and\ \bibinfo {author} {\bibfnamefont {J.}~\bibnamefont
  {Furthm{\"u}ller}},\ }\href@noop {} {\bibfield  {journal} {\bibinfo
  {journal} {Phys. Rev. B}\ }\textbf {\bibinfo {volume} {54}},\ \bibinfo
  {pages} {11169} (\bibinfo {year} {1996})}\BibitemShut {NoStop}%
\bibitem [{\citenamefont {Bl{\"o}chl}(1994)}]{blochl1994projector}%
  \BibitemOpen
  \bibfield  {author} {\bibinfo {author} {\bibfnamefont {P.~E.}\ \bibnamefont
  {Bl{\"o}chl}},\ }\href@noop {} {\bibfield  {journal} {\bibinfo  {journal}
  {Phys. Rev. B}\ }\textbf {\bibinfo {volume} {50}},\ \bibinfo {pages} {17953}
  (\bibinfo {year} {1994})}\BibitemShut {NoStop}%
\bibitem [{\citenamefont {Perdew}\ and\ \citenamefont
  {Wang}(1992)}]{perdew1992phys}%
  \BibitemOpen
  \bibfield  {author} {\bibinfo {author} {\bibfnamefont {J.}~\bibnamefont
  {Perdew}}\ and\ \bibinfo {author} {\bibfnamefont {Y.}~\bibnamefont {Wang}},\
  }\href@noop {} {\bibfield  {journal} {\bibinfo  {journal} {Phys. Rev. B}\
  }\textbf {\bibinfo {volume} {45}},\ \bibinfo {pages} {13244} (\bibinfo {year}
  {1992})}\BibitemShut {NoStop}%
\bibitem [{\citenamefont {Koelling}\ and\ \citenamefont
  {Harmon}(1977)}]{koelling1977technique}%
  \BibitemOpen
  \bibfield  {author} {\bibinfo {author} {\bibfnamefont {D.}~\bibnamefont
  {Koelling}}\ and\ \bibinfo {author} {\bibfnamefont {B.}~\bibnamefont
  {Harmon}},\ }\href@noop {} {\bibfield  {journal} {\bibinfo  {journal} {J.
  Phys. C:Solid State Phys.}\ }\textbf {\bibinfo {volume} {10}},\ \bibinfo
  {pages} {3107} (\bibinfo {year} {1977})}\BibitemShut {NoStop}%
\bibitem [{\citenamefont {Neugebauer}\ and\ \citenamefont
  {Scheffler}(1992)}]{neugebauer1992adsorbate}%
  \BibitemOpen
  \bibfield  {author} {\bibinfo {author} {\bibfnamefont {J.}~\bibnamefont
  {Neugebauer}}\ and\ \bibinfo {author} {\bibfnamefont {M.}~\bibnamefont
  {Scheffler}},\ }\href@noop {} {\bibfield  {journal} {\bibinfo  {journal}
  {Phys. Rev. B}\ }\textbf {\bibinfo {volume} {46}},\ \bibinfo {pages} {16067}
  (\bibinfo {year} {1992})}\BibitemShut {NoStop}%
\bibitem [{\citenamefont {Landau}\ and\ \citenamefont
  {Lifshitz}(1984)}]{principles}%
  \BibitemOpen
  \bibfield  {author} {\bibinfo {author} {\bibfnamefont {L.~D.}\ \bibnamefont
  {Landau}}\ and\ \bibinfo {author} {\bibfnamefont {E.~M.}\ \bibnamefont
  {Lifshitz}},\ }\href@noop {} {\emph {\bibinfo {title} {Electrodynamics of
  Continuous Media (Pergamon Press, Oxford,1984)}}}\ (\bibinfo  {publisher}
  {Pergamon Oxford},\ \bibinfo {year} {1984})\BibitemShut {NoStop}%
\bibitem [{\citenamefont {Shick}\ \emph {et~al.}(1997)\citenamefont {Shick},
  \citenamefont {Novikov},\ and\ \citenamefont
  {Freeman}}]{shick1997relativistic}%
  \BibitemOpen
  \bibfield  {author} {\bibinfo {author} {\bibfnamefont {A.}~\bibnamefont
  {Shick}}, \bibinfo {author} {\bibfnamefont {D.}~\bibnamefont {Novikov}}, \
  and\ \bibinfo {author} {\bibfnamefont {A.~J.}\ \bibnamefont {Freeman}},\
  }\href@noop {} {\bibfield  {journal} {\bibinfo  {journal} {Phys.Rev. B}\
  }\textbf {\bibinfo {volume} {56}},\ \bibinfo {pages} {R14259} (\bibinfo
  {year} {1997})}\BibitemShut {NoStop}%
\bibitem [{\citenamefont {Paes}\ and\ \citenamefont
  {Mosca}(2013)}]{paes2013effective}%
  \BibitemOpen
  \bibfield  {author} {\bibinfo {author} {\bibfnamefont {V.~Z.}\ \bibnamefont
  {Paes}}\ and\ \bibinfo {author} {\bibfnamefont {D.~H.}\ \bibnamefont
  {Mosca}},\ }\href@noop {} {\bibfield  {journal} {\bibinfo  {journal} {J.
  Magn. Magn. Mater.}\ }\textbf {\bibinfo {volume} {330}},\ \bibinfo {pages}
  {81} (\bibinfo {year} {2013})}\BibitemShut {NoStop}%
\bibitem [{\citenamefont {Komelj}\ and\ \citenamefont
  {F{\"a}hnle}(2000)}]{komelj2000magnetoelastic}%
  \BibitemOpen
  \bibfield  {author} {\bibinfo {author} {\bibfnamefont {M.}~\bibnamefont
  {Komelj}}\ and\ \bibinfo {author} {\bibfnamefont {M.}~\bibnamefont
  {F{\"a}hnle}},\ }\href@noop {} {\bibfield  {journal} {\bibinfo  {journal} {J.
  Magn. Magn. Mater.}\ }\textbf {\bibinfo {volume} {222}},\ \bibinfo {pages}
  {245} (\bibinfo {year} {2000})}\BibitemShut {NoStop}%
\bibitem [{\citenamefont {Tian}\ \emph {et~al.}(2009)\citenamefont {Tian},
  \citenamefont {Sander},\ and\ \citenamefont {Kirschner}}]{tian2009nonlinear}%
  \BibitemOpen
  \bibfield  {author} {\bibinfo {author} {\bibfnamefont {Z.}~\bibnamefont
  {Tian}}, \bibinfo {author} {\bibfnamefont {D.}~\bibnamefont {Sander}}, \ and\
  \bibinfo {author} {\bibfnamefont {J.}~\bibnamefont {Kirschner}},\ }\href@noop
  {} {\bibfield  {journal} {\bibinfo  {journal} {Phys. Rev. B}\ }\textbf
  {\bibinfo {volume} {79}},\ \bibinfo {pages} {024432} (\bibinfo {year}
  {2009})}\BibitemShut {NoStop}%
\bibitem [{\citenamefont {Ueda}\ \emph {et~al.}(2016)\citenamefont {Ueda},
  \citenamefont {Mizuguchi}, \citenamefont {Miura}, \citenamefont {Kang},
  \citenamefont {Shirai},\ and\ \citenamefont
  {Takanashi}}]{ueda2016electronic}%
  \BibitemOpen
  \bibfield  {author} {\bibinfo {author} {\bibfnamefont {S.}~\bibnamefont
  {Ueda}}, \bibinfo {author} {\bibfnamefont {M.}~\bibnamefont {Mizuguchi}},
  \bibinfo {author} {\bibfnamefont {Y.}~\bibnamefont {Miura}}, \bibinfo
  {author} {\bibfnamefont {J.}~\bibnamefont {Kang}}, \bibinfo {author}
  {\bibfnamefont {M.}~\bibnamefont {Shirai}}, \ and\ \bibinfo {author}
  {\bibfnamefont {K.}~\bibnamefont {Takanashi}},\ }\href@noop {} {\bibfield
  {journal} {\bibinfo  {journal} {Appl. Phys. Lett.}\ }\textbf {\bibinfo
  {volume} {109}},\ \bibinfo {pages} {042404} (\bibinfo {year}
  {2016})}\BibitemShut {NoStop}%
\bibitem [{\citenamefont {Ong}\ \emph {et~al.}(2015)\citenamefont {Ong},
  \citenamefont {Kioussis}, \citenamefont {Odkhuu}, \citenamefont {Amiri},
  \citenamefont {Wang},\ and\ \citenamefont {Carman}}]{ong2015giant}%
  \BibitemOpen
  \bibfield  {author} {\bibinfo {author} {\bibfnamefont {P.}~\bibnamefont
  {Ong}}, \bibinfo {author} {\bibfnamefont {N.}~\bibnamefont {Kioussis}},
  \bibinfo {author} {\bibfnamefont {D.}~\bibnamefont {Odkhuu}}, \bibinfo
  {author} {\bibfnamefont {P.~K.}\ \bibnamefont {Amiri}}, \bibinfo {author}
  {\bibfnamefont {K.}~\bibnamefont {Wang}}, \ and\ \bibinfo {author}
  {\bibfnamefont {G.~P.}\ \bibnamefont {Carman}},\ }\href@noop {} {\bibfield
  {journal} {\bibinfo  {journal} {Phys. Rev. B}\ }\textbf {\bibinfo {volume}
  {92}},\ \bibinfo {pages} {020407} (\bibinfo {year} {2015})}\BibitemShut
  {NoStop}%
\bibitem [{\citenamefont {Wang}\ \emph {et~al.}(1993)\citenamefont {Wang},
  \citenamefont {Wu},\ and\ \citenamefont {Freeman}}]{wang1993first}%
  \BibitemOpen
  \bibfield  {author} {\bibinfo {author} {\bibfnamefont {D.~S.}\ \bibnamefont
  {Wang}}, \bibinfo {author} {\bibfnamefont {R.}~\bibnamefont {Wu}}, \ and\
  \bibinfo {author} {\bibfnamefont {A.~J.}\ \bibnamefont {Freeman}},\
  }\href@noop {} {\bibfield  {journal} {\bibinfo  {journal} {Phys. Rev. B}\
  }\textbf {\bibinfo {volume} {47}},\ \bibinfo {pages} {14932} (\bibinfo {year}
  {1993})}\BibitemShut {NoStop}%
\bibitem [{\citenamefont {Nakamura}\ \emph {et~al.}(2010)\citenamefont
  {Nakamura}, \citenamefont {Akiyama}, \citenamefont {Ito}, \citenamefont
  {Weinert},\ and\ \citenamefont {Freeman}}]{kohji10:prb}%
  \BibitemOpen
  \bibfield  {author} {\bibinfo {author} {\bibfnamefont {K.}~\bibnamefont
  {Nakamura}}, \bibinfo {author} {\bibfnamefont {T.}~\bibnamefont {Akiyama}},
  \bibinfo {author} {\bibfnamefont {T.}~\bibnamefont {Ito}}, \bibinfo {author}
  {\bibfnamefont {M.}~\bibnamefont {Weinert}}, \ and\ \bibinfo {author}
  {\bibfnamefont {A.~J.}\ \bibnamefont {Freeman}},\ }\href@noop {} {\bibfield
  {journal} {\bibinfo  {journal} {Phys. Rev. B}\ }\textbf {\bibinfo {volume}
  {81}},\ \bibinfo {pages} {220409(R)} (\bibinfo {year} {2010})}\BibitemShut
  {NoStop}%
\bibitem [{\citenamefont {Odkhuu}\ \emph {et~al.}(2011)\citenamefont {Odkhuu},
  \citenamefont {Yun}, \citenamefont {Rhim},\ and\ \citenamefont
  {Hong}}]{odkhuu2011engineering}%
  \BibitemOpen
  \bibfield  {author} {\bibinfo {author} {\bibfnamefont {D.}~\bibnamefont
  {Odkhuu}}, \bibinfo {author} {\bibfnamefont {W.~S.}\ \bibnamefont {Yun}},
  \bibinfo {author} {\bibfnamefont {S.~H.}\ \bibnamefont {Rhim}}, \ and\
  \bibinfo {author} {\bibfnamefont {S.~C.}\ \bibnamefont {Hong}},\ }\href@noop
  {} {\bibfield  {journal} {\bibinfo  {journal} {Appl. Phys. Lett.}\ }\textbf
  {\bibinfo {volume} {98}},\ \bibinfo {pages} {152502} (\bibinfo {year}
  {2011})}\BibitemShut {NoStop}%
\bibitem [{\citenamefont {Odkhuu}\ \emph {et~al.}(2013)\citenamefont {Odkhuu},
  \citenamefont {Rhim}, \citenamefont {Park},\ and\ \citenamefont
  {Hong}}]{odkhuu13:prb}%
  \BibitemOpen
  \bibfield  {author} {\bibinfo {author} {\bibfnamefont {D.}~\bibnamefont
  {Odkhuu}}, \bibinfo {author} {\bibfnamefont {S.~H.}\ \bibnamefont {Rhim}},
  \bibinfo {author} {\bibfnamefont {N.}~\bibnamefont {Park}}, \ and\ \bibinfo
  {author} {\bibfnamefont {S.~C.}\ \bibnamefont {Hong}},\ }\href@noop {}
  {\bibfield  {journal} {\bibinfo  {journal} {Phys. Rev. B}\ }\textbf {\bibinfo
  {volume} {88}},\ \bibinfo {pages} {184405} (\bibinfo {year}
  {2013})}\BibitemShut {NoStop}%
\bibitem [{\citenamefont {Hotta}\ \emph {et~al.}(2013)\citenamefont {Hotta},
  \citenamefont {Nakamura}, \citenamefont {Akiyama}, \citenamefont {Ito},
  \citenamefont {Oguchi},\ and\ \citenamefont {Freeman}}]{hotta13:prl}%
  \BibitemOpen
  \bibfield  {author} {\bibinfo {author} {\bibfnamefont {K.}~\bibnamefont
  {Hotta}}, \bibinfo {author} {\bibfnamefont {K.}~\bibnamefont {Nakamura}},
  \bibinfo {author} {\bibfnamefont {T.}~\bibnamefont {Akiyama}}, \bibinfo
  {author} {\bibfnamefont {T.}~\bibnamefont {Ito}}, \bibinfo {author}
  {\bibfnamefont {T.}~\bibnamefont {Oguchi}}, \ and\ \bibinfo {author}
  {\bibfnamefont {A.~J.}\ \bibnamefont {Freeman}},\ }\href@noop {} {\bibfield
  {journal} {\bibinfo  {journal} {Phys. Rev. Lett.}\ }\textbf {\bibinfo
  {volume} {110}},\ \bibinfo {pages} {267206} (\bibinfo {year}
  {2013})}\BibitemShut {NoStop}%
\bibitem [{\citenamefont {Odkhuu}\ \emph {et~al.}(2016)\citenamefont {Odkhuu},
  \citenamefont {Yun}, \citenamefont {Rhim},\ and\ \citenamefont
  {Hong}}]{odkhuu2016126}%
  \BibitemOpen
  \bibfield  {author} {\bibinfo {author} {\bibfnamefont {D.}~\bibnamefont
  {Odkhuu}}, \bibinfo {author} {\bibfnamefont {W.~S.}\ \bibnamefont {Yun}},
  \bibinfo {author} {\bibfnamefont {S.~H.}\ \bibnamefont {Rhim}}, \ and\
  \bibinfo {author} {\bibfnamefont {S.~C.}\ \bibnamefont {Hong}},\ }\href
  {\doibase https://doi.org/10.1016/j.jmmm.2016.04.027} {\bibfield  {journal}
  {\bibinfo  {journal} {J. Magn. Magn. Mater.}\ }\textbf {\bibinfo {volume}
  {414}},\ \bibinfo {pages} {126 } (\bibinfo {year} {2016})}\BibitemShut
  {NoStop}%
\bibitem [{\citenamefont {{Qurat-ul-ain}}\ \emph {et~al.}(2018)\citenamefont
  {{Qurat-ul-ain}}, \citenamefont {Cuong}, \citenamefont {Odkhuu},
  \citenamefont {Rhim},\ and\ \citenamefont {Hong}}]{qurat201869}%
  \BibitemOpen
  \bibfield  {author} {\bibinfo {author} {\bibnamefont {{Qurat-ul-ain}}},
  \bibinfo {author} {\bibfnamefont {D.~D.}\ \bibnamefont {Cuong}}, \bibinfo
  {author} {\bibfnamefont {D.}~\bibnamefont {Odkhuu}}, \bibinfo {author}
  {\bibfnamefont {S.~H.}\ \bibnamefont {Rhim}}, \ and\ \bibinfo {author}
  {\bibfnamefont {S.~C.}\ \bibnamefont {Hong}},\ }\href {\doibase
  https://doi.org/10.1016/j.jmmm.2018.07.055} {\bibfield  {journal} {\bibinfo
  {journal} {J. Magn. Magn. Mater.}\ }\textbf {\bibinfo {volume} {467}},\
  \bibinfo {pages} {69 } (\bibinfo {year} {2018})}\BibitemShut {NoStop}%
\end{thebibliography}
%
\end{document}


\preprint{ver.20 / \today}
\title{Supplementary Material:\\ \title{Enhanced voltage-controlled magnetic anisotropy via magneto-elasticity in FePt/MgO(001)}}

\author{Qurat-ul-ain$^{1}$}

\author{D. Odkhuu$^{2}$}

\author{S. H. Rhim$^{1}$}
\email[Email address: ]{sonny@ulsan.ac.kr}

\author{S. C. Hong$^{1}$}%
 \email[Email address: ]{schong@ulsan.ac.kr}
\affiliation{$^{1}$Department of Physics and Energy Harvest-Storage Research Center, University of Ulsan, Ulsan, Republic of Korea\\
$^{2}$Department of Physics, Incheon National University, Incheon, Republic of Korea
}%

\date{\today}
\maketitle

\section{Derivation of magneto-crystalline anisotropy energy density }

The magneto-crystalline anisotropy energy ($E_{MA}$) is written as
\begin{align}
E_{MA}=K_1t(1-\alpha_3^2)+b_1 t \sum_{k=1}^{3}\eta_k\alpha_k^2
        +\frac{1}{2}B_1 t\sum_{k=1}^{3}\eta_k^2\alpha_k^2,\label{eq:MA1_S}
\end{align}
where the first-term is the zero strain anisotropy energy and the latter two are the magneto-elastic energy terms\cite{eastman1966ultrasonic}. 
As mentioned in the main text, $K_1$ is the anisotropy constant,
$\alpha_{k}(k=1,2,3)$ are the direction cosines of magnetization with respect to the principal axis,
$b_1$ and $B_1$ are the first- and second-order
magneto-elastic coefficients, respectively\cite{principles}, and $\eta_{k}$ ($k=1-6$) is the strain tensor in Voigt's notation.
For a thin film grown epitaxially along the [001] crystallographic directions, shear strains, i.e., $\eta_4=\eta_5=\eta_6=0$
and $\eta_1=\eta_2=\eta$ due to tetragonal symmetry.
The lattice strain $\eta_k$ in Eq. \ref{eq:MA1_S} can be calculated via boundary conditions of the problem
using the magneto-elastic equation of state, given as:
\begin{align}
\sigma_{k}=\frac{\partial}{\partial \eta_k}\left(b_1 t\sum_{k=1}^{3}\eta_k\alpha_k^2+\frac{1}{2}B_1 t \sum_{k=1}^{3}\eta_k^2\alpha_k^2
             +\frac{1}{2}c_{11}\sum_{k=1}^{3}\eta_k^2+c_{12}\sum_{\substack{j,k=1\\j\neq k}}^3\eta_j\eta_k\right) 
             \label{eq:MA2_S}
\end{align} 
where $\sigma_k$ is the mechanical stress.
$c_{11}$ and $c_{12}$ are the elastic stiffnesses at constant magnetization.
From the condition $\sigma_3=0$, which follows from the relaxation of the interlayer distance
perpendicular to the film plane.
One further obtains:
\begin{align}
\eta_3=-\frac{2\eta c_{12}+b_1\alpha_3^2}{c_{11}+B_1\alpha_3^2},
\end{align}

From the Poisson's ratio expression $\nu=c_{12}/(c_{11}+c_{12})$, we obtain $c_{12}/c_{11}=\nu/(1-\nu)$.
Where $\nu\approx\frac{1}{3}$ for transition metals\cite{sander1999correlation}, which gives $c_{12}=c_{11}/2$.
The zero strain anisotropy energy can be further decomposed into bulk and interface contributions, 
$K_1 = K_1^v + K_1^i/t \approx K_1^i/t$ for thin film limit.
By substituting $\eta_3$ in Eq. \ref{eq:MA1_S}, and calculating the energy required to rotate magnetization from in-plane 
($\alpha_1=1$) to out-of-plane ($\alpha_3=1$), we get:
\begin{eqnarray}
\label{eq:MA2}
  E_{MA} = K_{eff}
          + \left( 1+\omega\right)b_1t\eta 
          + \left( 1-\omega\right)\frac{B_1}{2}t\eta^2,\label{eq:MA_S}
\end{eqnarray}
where
\begin{eqnarray}
\label{eq:MA3}
  K_{eff} = K_1^i+\omega\frac{b_1^2}{c_{11}}\left(1+\frac{B_1}{2c_{11}}\right)t,  
\end{eqnarray}
and
\begin{eqnarray}
\label{eq:MA3}
  \omega = c_{11}^2/\left( c_{11}+B_1\right)^2.  
\end{eqnarray}

\section{Orbital resolved band structure}
\begin{figure}[!htbp]
\centering
\hspace{1cm}
\includegraphics[width=\columnwidth]{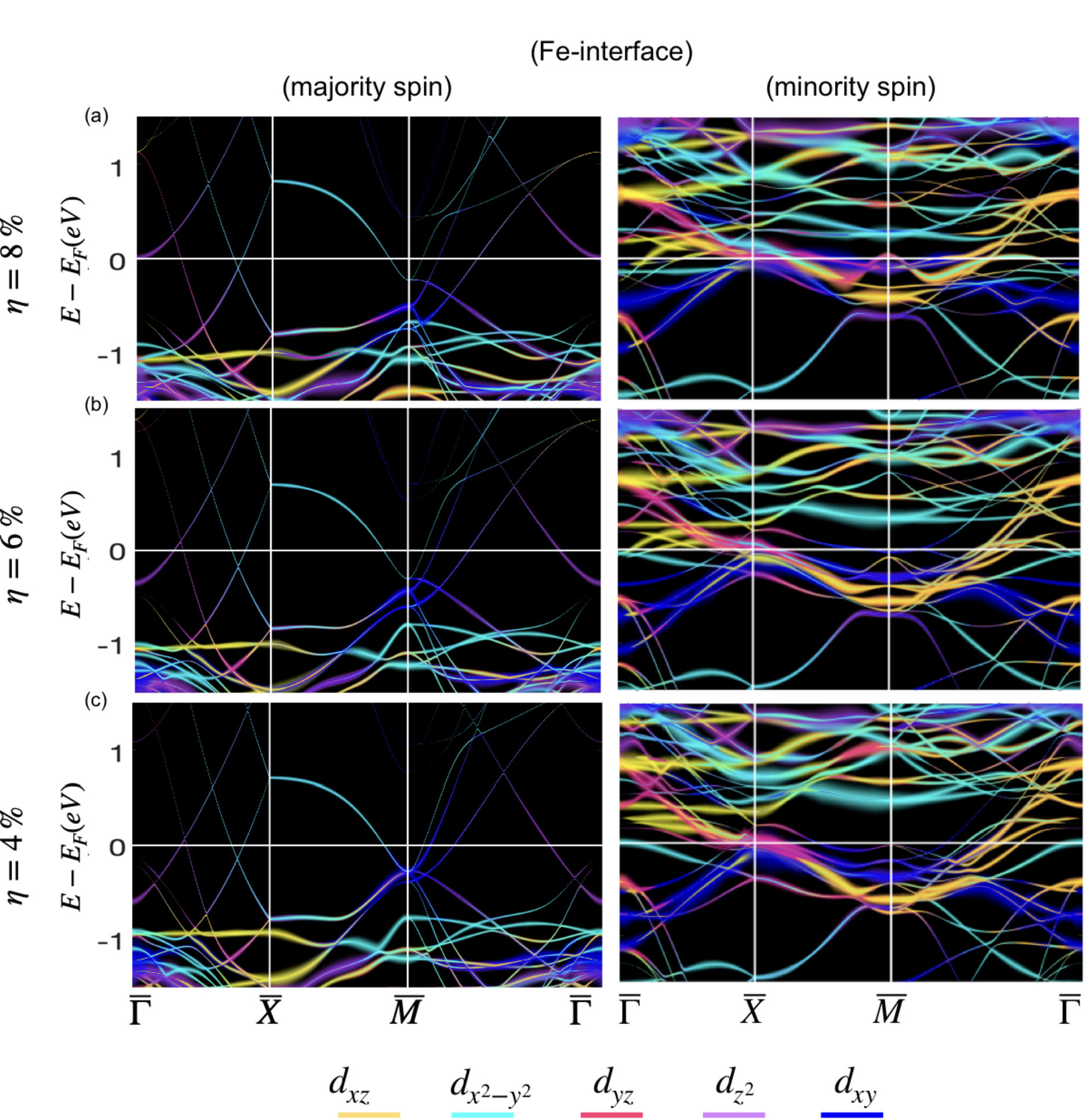}
\caption{
Orbitally resolved Fe {\em d} bands for majority and minority spins along $\overline\Gamma$-$\overline {X}$-$\overline {M}$-$\overline {\Gamma}$ direction 
in 2D Brilliouin zone at the FM/I interface for $\eta = 8, 6$, and $4\%$, respectively.
The reference energy (E= 0 eV) places at $E_F$. 
Blue, cyan, pink, purple and golden color for {\em d$_{xy}$}, {\em d$_{x^2-y^2}$},
{\em d$_{yz}$}, {\em d$_{z^2}$} and {\em d$_{xz}$}.
  }
\label{fig:S1}
\end{figure}

\begin{figure}[!htbp]
\centering
\hspace{1cm}
\includegraphics[width=\columnwidth]{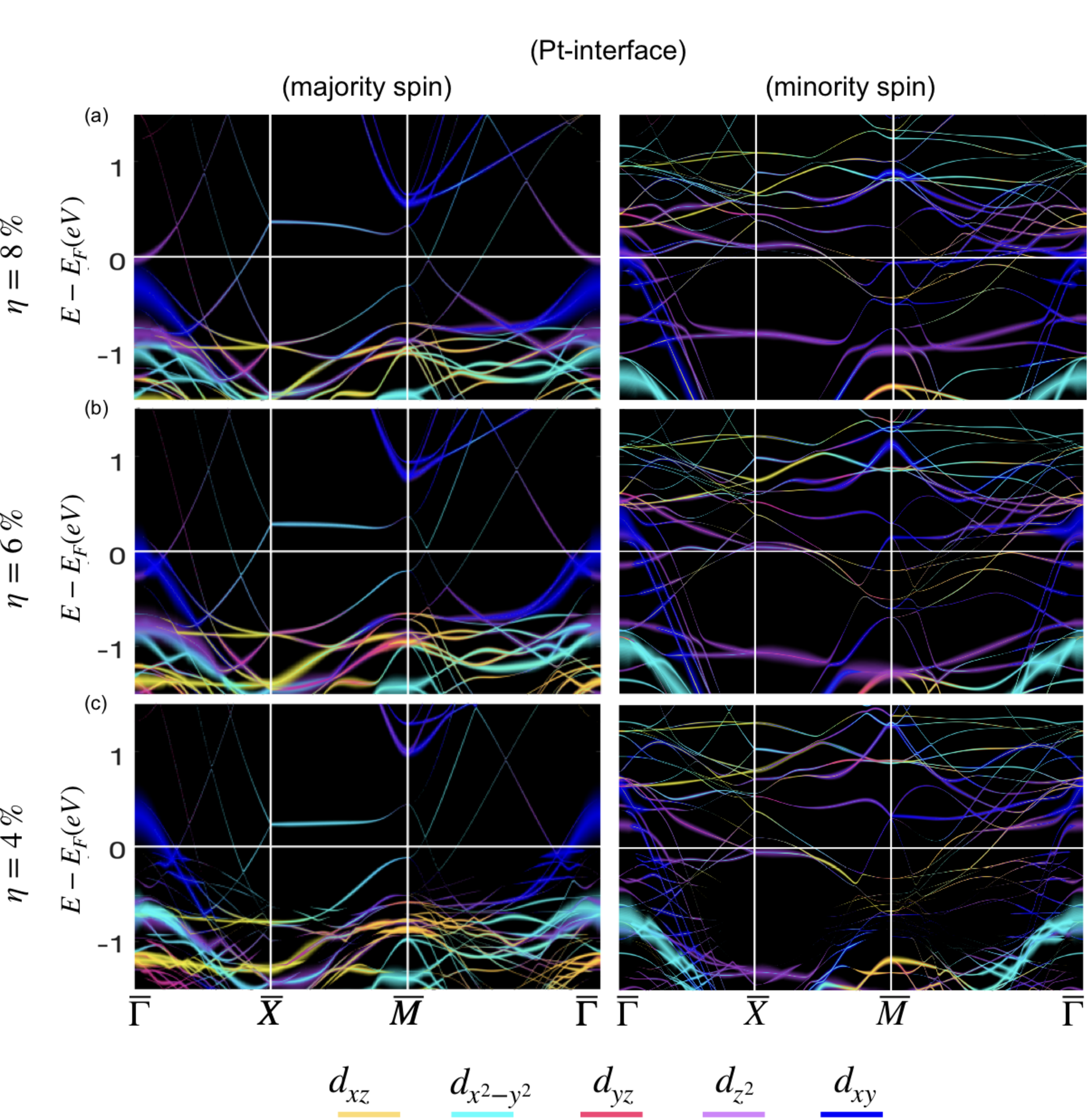}
\caption{
Orbitally resolved Pt {\em d} bands for majority and minority spins,
along $\overline\Gamma$-$\overline {X}$-$\overline {M}$-$\overline {\Gamma}$ direction 
in 2D Brilliouin zone at the FM/I interface for $\eta = 8, 6$, and $4\%$, respectively.
The reference energy (E= 0 eV) places at $E_F$. 
Blue, cyan, pink, purple and golden color for {\em d$_{xy}$}, {\em d$_{x^2-y^2}$},
{\em d$_{yz}$}, {\em d$_{z^2}$} and {\em d$_{xz}$}.
  }
\label{fig:S2}
\end{figure}

Orbital resolved bands at $\eta=$ 4\%, 6\%, and 8\% are plotted along high symmetry direction in 2-D Brillouin zone as shown in Fig.~\ref{fig:S1} for Fe-interface and Fig.~\ref{fig:S2} for Pt-interface. 
For Fe-interface, majority spins are almost occupied.
Most of the perpendicular magneto-crystalline anisotropy (PMA) arises from minority spin channel.
For Pt-interface, minority spins are the major contributor to PMA.
With increasing compressive strain, Fe and Pt {\em d} bands overall shift towards lower energy.

\section{Spin-orbit coupling between {\em d} bands}
In Table.~\ref{tab:matrix_full}, we provide a list of positive spin-orbit coupling (SOC) between {\em d} bands for both Fe- and Pt-interfaces. 
The strength of the SOC is $\Delta^{\alpha}=\frac{1}{e_u - e_o}$.
Here, $e_u$ and $e_o$ are the energies of unoccupied and occupied bands;
$\alpha=+,0,-$ denotes when $E_{eff} > 0 $, $E_{eff}=0$, and $E_{eff}<0$, respectively.
For Fe-interface, {\em d$_{xy}$} and {\em d$_{x^2-y^2}$} bands couples at $\overline X$ providing large PMA at $E_{eff}=0$. 
Under $E_{eff}>0$, this coupling increased up to two times due to band shift towards $E_F$.
Alternatively, due to band shift away from $E_F$ under $E_{eff}<0$, this coupling reduces.
Moreover, {\em d$_{xz}$} and {\em d$_{yz}$} bands also yield large PMA at $\overline M$.
For Pt-interface, electric field effects the SOC strength linearly, i.e. $\Delta^-> \Delta^0> \Delta^+$.
{\em d$_{xy}$} and {\em d$_{x^2-y^2}$} bands at $\overline X$ shift towards 
$E_F$ under $E_{eff} <0$, providing large PMA.
However, these bands move away from $E_F$ under $E_{eff} >0$, reducing PMA contribution.

\begin{table}[ht!]
  \centering
\setlength{\extrarowheight}{2pt} 
\ra{1.0}
\begin{tabular}{@{}lr   lr   cr   cr   cr   cr   c@{}}\toprule[0.5mm]

&&  symmetry point  &&      coupled bands             &&    && $\Delta^-$ && $\Delta^0$ & & $\Delta^+$ \\[1ex]
 \midrule[0.1mm]
  \midrule[0.1mm]
Fe-interface& & $\overline X$         && $\textit{d}_{xy}$, $\textit{d}_{x^2-y^2}$  &&  && 2.558   &&  6.897  &&   12.300    \\
  &&      &&  $\textit{d}_{xy}$, $\textit{d}_{x^2-y^2}$ &&  && 2.004  &&  2.151  &&   2.015    \\[-0.5in]
  &&           && $\textit{d}_{xz}$,$\textit{d}_{yz}$ &&  && 5.540   &&  5.903  &&   6.763    \\
   &&           && $\textit{d}_{xz}$,$\textit{d}_{yz}$ &&   && 1.998  &&  2.010  &&   2.109    \\[1ex]  
   \cline{3-13}
&&   $\overline M$         && $\textit{d}_{xy}$, $\textit{d}_{x^2-y^2}$ && && 2.782  &&  2.782  &&   2.961    \\
 &&             && $\textit{d}_{xy}$, $\textit{d}_{x^2-y^2}$ &&   && 1.530  &&  1.487  &&   1.541  \\[-0.5in] 
  &&           && $\textit{d}_{xz}$,$\textit{d}_{yz}$ &&  && 1.024   &&  1.011  &&   1.075    \\[0.2in] 
   \cline{3-13}
  && $\frac{1}{2}\overline {XM}$  && $\textit{d}_{xy}$, $\textit{d}_{x^2-y^2}$ &&  && 1.783  &&  1.700  &&   1.792    \\
   &&           && $\textit{d}_{xy}$, $\textit{d}_{x^2-y^2}$ &&  && 1.447  &&  1.335  &&   1.475  \\[-0.5in] 
    &&         && $\textit{d}_{x^2-y^2}$, $\textit{d}_{xy}$ &&  && 0.000   &&  7.100  &&   0.000    \\
      &&        && $\textit{d}_{xz}$, $\textit{d}_{yz}$ &&   && 1.024   &&  1.011  &&  1.065    \\[1ex]
   \midrule[0.1mm]
   \midrule[0.1mm]
Pt-interface& & $\overline X$         && $\textit{d}_{xy}$, $\textit{d}_{x^2-y^2}$  &&   && 0.766  &&  0.769  &&   0.760    \\
   \cline{3-13}
&&   $\overline M$         && $\textit{d}_{x^2-y^2}$, $\textit{d}_{xy}$ &&   &&0.643  && 0.640  &&  0.638    \\
   \cline{3-13}
  && $\frac{1}{2}\overline {M\Gamma}$  && $\textit{d}_{x^2-y^2}$, $\textit{d}_{xy}$ &&   && 0.526 &&  0.524  && 0.524    \\
 \bottomrule[0.5mm]
\end{tabular}
\caption{ List of SOC strength between different {\em d} bands at high symmetry points in 2D BZ for both Fe- and Pt-interface.
$\Delta^{\alpha}=\frac{1}{e_u - e_o}$ represents the strength of SOC,
where $e_u$ ($e_o$) are energies of unoccupied (occupied) band;
$\alpha=+,0,-$ denotes when $E_{eff} > 0 $, $E_{eff}=0$, and $E_{eff}<0$, respectively.} 
\label{tab:matrix_full}
\end{table}

%